\documentclass[cameraready]{Interspeech}

\title{Multi-Source Evidence Fusion for Audio Question Answering}

\author{Aivo}{Olev}
\author{Tanel}{Alumäe}

\address{
    Tallinn University of Technology, Estonia
}

\email{\{aivo.olev,tanel.alumae\}@taltech.ee}

\keywords{audio reasoning, reasoning quality, large audio language models, evidence combination, tool reliability}

\usepackage{comment}
\usepackage{tikz}
\usetikzlibrary{shapes.geometric, arrows.meta, positioning, fit, calc, backgrounds}
\usepackage{booktabs}
\usepackage{multirow}
\usepackage{xcolor}
\usepackage{enumitem}
\usepackage{pifont}
\usepackage{hyperref}

\setlist{nosep,leftmargin=*}

\begin{document}
\maketitle

% ===================================================================================
% ABSTRACT (max 1000 characters, ASCII only, no citations)
% ===================================================================================
\begin{abstract}
Large audio language models (LALMs) can answer questions about speech, music, and environmental sounds, yet their internal reasoning is largely opaque and difficult to validate. We describe TalTech's solution to the Agent Track of the Interspeech 2026 Audio Reasoning Challenge, in which systems are evaluated on reasoning process quality, specifically the factual accuracy, logical soundness, and completeness of their reasoning chains. Our multi-source ensemble pipeline uses two LALMs that generate independent observations, while a separate text-only reasoning model cross-checks these against outputs from 25 acoustic tools organized into reliability tiers. By grounding every inference step in explicit, reliability-tagged evidence, the system produces dense, verifiable reasoning chains. Our system ranked first in the challenge, outperforming all competing systems by a wide margin in  challenge’s reasoning quality metric. %Ablation studies show that combining evidence from dual sources is the primary contributor to the accuracy of our system.
\end{abstract}

% ===================================================================================
% 1. INTRODUCTION
% ===================================================================================
\section{Introduction}
\label{sec:intro}

Audio understanding has advanced rapidly with the emergence of large audio language models (LALMs)~\cite{chu2023qwenaudio, tang2024salmonn, kong2024audioflamingo, gong2024ltu}.
Yet these models often produce opaque reasoning that resists verification. This limitation is masked by benchmarks that evaluate only final-answer accuracy~\cite{ma2026audioreasoningchallenge}.
The Interspeech 2026 Audio Reasoning Challenge~\cite{audioreasoningchallenge2026, ma2026audioreasoningchallenge, ma2025mmar} addresses this gap: its tasks require combining perception with multi-step reasoning across music theory, speaker analysis, scene understanding, and temporal reasoning, and systems are ranked primarily on reasoning process quality (factuality, logic, and completeness of reasoning chains), not accuracy alone.
No single model excels at all of these capabilities, motivating agent-based systems that orchestrate multiple specialized tools.
This paper describes our entry to the Agent Track, focusing on how we combine heterogeneous audio tools and speech LLMs to produce both accurate and transparently reasoned answers.

A central challenge for such systems is that their information sources have fundamentally different reliability characteristics.
LALMs provide rich, high-level observations but are prone to hallucination: they may fabricate timestamps, report visual information from audio-only input, or overcount events.
Traditional acoustic tools (e.g., beat detection, spectral analysis) produce reproducible measurements but can only answer narrow questions and often struggle with real-world data that might contain noise, different types of sounds and out-of-domain audio events.
Automatic speech recognition occupies a middle ground: it is generally reliable but subject to errors.
Therefore, it is difficult to build an agent that reasons correctly when combining evidence from sources that span this reliability spectrum. Ensemble methods assume roughly equal participant reliability~\cite{wang2025moa, du2024debate}, and agentic frameworks treat tools as equally reliable oracles~\cite{yao2023react, schick2023toolformer}.

Our system incorporates three design decisions that proved effective in the challenge setting:
\begin{enumerate}
    \item \textbf{Dual-source evidence fusion} in which two speech LLMs independently report observations and a downstream reasoning model cross-validates them.
    \item A \textbf{four-tier tool reliability framework} with confidence caps, relevance scoring, corroboration bonuses, and domain appropriateness adjustments for 25 tools.
    \item A \textbf{three-stage contradiction detection} mechanism with hypothesis-driven targeted verification.
\end{enumerate}

The system ranked first in the Agent Track according to the challenge’s primary metric, reasoning quality (MMAR-Rubrics: 69.8), and achieved the second-highest accuracy (76.9\%). %The margin in reasoning quality over the second-place system was 3.6~pp.
Our evidence-based architecture encourages dense and verifiable reasoning. Each claim is supported by reliability-tagged observations, tool measurements, or corroboration across sources, producing reasoning chains with many checkable factual statements (Section~\ref{sec:discussion}).
Ablation experiments show that dual-source evidence fusion yields a statistically significant improvement in accuracy.

While the reliability weights and confidence caps were tuned for this specific challenge, the underlying architectural principles of evidence separation, tiered reliability, and contradiction driven verification may generalize to other reasoning tasks that involve heterogeneous sources.

% ===================================================================================
% 2. RELATED WORK
% ===================================================================================
\section{Related work}
\label{sec:related}

LALMs have advanced rapidly~\cite{chu2023qwenaudio, chu2024qwen2audio, xu2025qwen3omni, tian2025stepaudior1}, yet persistent hallucination across 14 distinct failure modes~\cite{cheng2025ahabench} and strong affirmation bias~\cite{kuan2024hallucination} limit single-model reliability.
Reinforcement learning improves reasoning but not audio perception itself~\cite{rouditchenko2025omnir1}, motivating multi-source approaches.

Agentic frameworks such as ReAct~\cite{yao2023react} and Toolformer~\cite{schick2023toolformer} treat tools as equally reliable oracles.
Audio-specific agents, such as AudioGPT~\cite{huang2024audiogpt}, AudioToolAgent~\cite{wijngaard2025audiotoolagent}, and AudioGenie-Reasoner~\cite{rong2025audiogenie}, coordinate multiple models but lack explicit reliability hierarchies or confidence caps.
Sun et al.~\cite{sun2024toolsfail} showed that tools can produce plausible but incorrect results without error signals, motivating reliability-weighted combination.

Multi-model ensemble methods, such as Mixture-of-Agents~\cite{wang2025moa} and multi-agent debate~\cite{du2024debate}, assume roughly homogeneous participant reliability.
Hwang et al.~\cite{hwang2024rarag} demonstrated that explicit reliability modeling improves robustness under heterogeneous source quality.
Our pipeline hides LALM answer predictions from the reasoning agent as a precaution against anchoring on potentially incorrect predictions, though our ablation suggests the effect is small in this setting (Section~\ref{sec:ablation_results}).

Sycophancy and anchoring bias cause LLMs to produce unfaithful reasoning when exposed to suggested answers~\cite{turpin2023unfaithful, sharma2024sycophancy}.
%Whisper assigns high confidence to 10--20\% of incorrect tokens under noise~\cite{huo2025whisper}, motivating our confidence caps.
Our evidence-only design hides LALM final answer predictions from the reasoning agent to prevent anchoring on potentially incorrect predictions.

% ===================================================================================
% 3. SYSTEM ARCHITECTURE
% ===================================================================================
\section{System architecture}
\label{sec:system}

%\subsection{Overview}
\label{sec:overview}

\begin{figure*}[t]
\centering
\resizebox{\textwidth}{!}{%
\begin{tikzpicture}[
    node distance=0.3cm and 0.35cm,
    >={Stealth[length=3pt]},
    box/.style={draw, rounded corners=2pt, minimum height=0.6cm, text width=1.7cm, align=center, font=\scriptsize},
    smallbox/.style={draw, rounded corners=2pt, minimum height=0.55cm, text width=1.4cm, align=center, font=\scriptsize},
]

% Input
\node[box, fill=black!8] (input) {Audio + Question\\+ Choices};

% Speech LLMs (parallel) — positioned relative to a midpoint
\node[smallbox, fill=white, right=0.4cm of input, yshift=0.35cm] (slm1) {StepAudioR1\\{\tiny(full + 3 seg.)}};
\node[smallbox, fill=white, right=0.4cm of input, yshift=-0.35cm] (slm2) {Qwen3-Omni\\{\tiny(full + 3 seg.)}};

% Unified analysis — positioned to the right of the speech LLM boxes
\node[box, fill=black!8, right=0.35cm of slm1.east, anchor=west, yshift=-0.35cm] (unified) {Unified Evidence\\Analysis};

% Tool loop - Step 1
\node[box, fill=black!8, right=0.35cm of unified] (step1) {Step 1: Evidence\\{\tiny(12 [23] tools)}};

% Contradiction detection
\node[box, fill=white, right=0.3cm of step1] (contra) {Contradiction\\Detection};

% Tool loop - Step 2
\node[box, fill=black!8, right=0.3cm of contra] (step2) {Step 2: Validate\\{\tiny(5 [8] tools)}};

% Answer Selection + Reasoning (combined)
\node[box, fill=black!8, right=0.3cm of step2] (anssel) {Answer\\Selection};

\node[box, fill=black!8, right=0.3cm of anssel] (reason) {Answer +\\Reasoning};

% Forward arrows
\draw[->] (input) -- (slm1);
\draw[->] (input) -- (slm2);
\draw[->] (slm1.east) -- ++(0.15,0) |- (unified);
\draw[->] (slm2.east) -- ++(0.15,0) |- (unified);
\draw[->] (unified) -- (step1);
\draw[->] (step1) -- (contra);
\draw[->] (contra) -- (step2);
\draw[->] (step2) -- (anssel);
\draw[->] (anssel) -- (reason);

% Self-loop arrows for iterative steps
\draw[->, thick] (step1.north) to[out=60, in=120, looseness=3.5] node[above, font=\scriptsize] {$\leq$3$\times$} (step1.north);
\draw[->, thick] (step2.north) to[out=60, in=120, looseness=3.5] node[above, font=\scriptsize] {$\leq$2$\times$} (step2.north);

\end{tikzpicture}%
}% end resizebox
\caption{Multi-source ensemble pipeline. Two speech LLMs analyze audio across four segments each; unified analysis identifies corroborations and disagreements. Step~1 gathers evidence from reliability-tiered tools (up to 3 rounds), three-stage contradiction detection generates verification hypotheses, and Step~2 performs targeted segment-level validation. Bracket counts indicate additional music-specific tools.}
\label{fig:architecture}
\end{figure*}
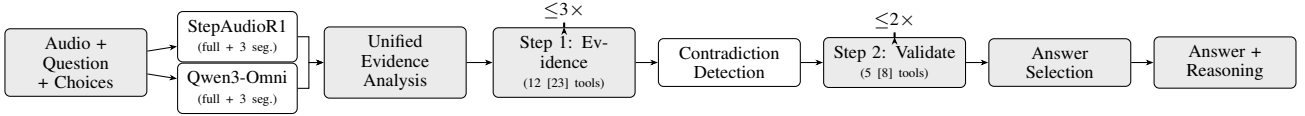

The system takes an audio file, a question, and multiple choice options as input and produces a selected answer with accompanying reasoning. Figure~\ref{fig:architecture} illustrates the pipeline. Two LALMs independently report observations about the audio. The reasoning agent receives these observations together with tool outputs tagged with confidence and relevance scores, but does not see the source models' answer predictions.

% --- ARCHITECTURE FIGURE ---

%\subsection{Multi-segment speech LLM evidence collection}
\label{sec:multiseg}

Two open-weights LALMs --- StepAudioR1~\cite{tian2025stepaudior1} and Qwen3-Omni~\cite{xu2025qwen3omni} --- independently analyze the audio.
Each model receives four queries: one for the full audio and one for each of three equal-duration segments, enabling focused temporal analysis.
Models are prompted to report observations, not to select an answer.
The synthesis step merges observations across segments, applies corroboration bonuses when segment and full-audio analyses agree, and classifies the audio content type.

%\subsection{Unified evidence analysis}
\label{sec:unified}

A reasoning LLM call performs semantic corroboration of both speech LLM outputs, classifying each observation as \emph{corroborated} (both sources agree; confidence 0.80--0.95), \emph{source-specific} (one source only; 0.50--0.70), or \emph{disagreement} (conflicting claims with credibility assessment).

%\subsection{Tool reliability framework}
\label{sec:reliability}

We classify each tool into four reliability tiers based on output determinism and reproducibility (Table~\ref{tab:reliability}).

\begin{table}[t]
\caption{Tool reliability tiers with representative tools, default confidence caps, and evidence weights.}
\label{tab:reliability}
\centering
\small
\addtolength{\tabcolsep}{-0.2em}
\begin{tabular}{@{}lp{3.9cm}cc@{}}
\toprule
\textbf{Tier} & \textbf{Representative tools} & \textbf{Cap} & \textbf{Weight} \\
\midrule
Analytic & Beat detection, energy dynamics, spectral features & 0.90 & 1.0 \\
Probabilistic & Whisper ASR, diarization, source separation, instruments & 0.75 & 0.75 \\
Heuristic     & Chord analysis, environment detection, scene context & 0.60 & 0.50 \\
LALMs    & StepAudioR1, Qwen3-Omni & 0.70 & 0.40 \\
\bottomrule
\end{tabular}
\end{table}

Each evidence item receives a confidence score combining the tier's base confidence, a $1.5\times$ corroboration multiplier, and a $1.3\times$ direct-answer bonus.
LALM evidence is capped at 0.70 regardless of bonuses. All caps, weights, and multipliers were set empirically during challenge development rather than learned from data.
Confidence and relevance are tracked independently; the argumentation stage weights evidence by their product.
Tools applied outside their primary domain (e.g., beat detection on speech) receive a reduced confidence via a domain appropriateness factor.

%\subsection{Tool verification and contradiction detection}
\label{sec:toolloop}

When the unified analysis reveals disagreements or insufficient confidence, the system enters a two-step verification loop.
In Step~1, the reasoning agent iteratively selects from 12 whole-audio tools (23 for music content).
After gathering evidence, a three-stage contradiction detector processes the combined evidence:
(1)~heuristic reclassification adjusts LALM confidence based on keyword overlap with tool outputs ($\pm 0.15$ for reproducible tools);
(2)~hallucination risk assessment assigns per-item risk levels based on reliability tier, corroboration status, and domain-specific guards (e.g., cosine-clustering speaker counts $\geq 3{\times}$ the diarization estimate are marked as segmentation artifacts);
(3)~an LLM contradiction detector classifies inter-tool conflicts, intra-tool inconsistencies, and reliability hierarchy violations, while checking four logical pitfalls: treating absence of detection as proof of absence, dismissing single-source claims instead of marking them speculative, diarization over-segmentation, and transcription disagreements where segment timings do not actually overlap.
A non-dismissal policy keeps LALM claims flagged as speculative rather than hallucinated unless actively contradicted by reproducible tool evidence, preserving potentially valid observations for downstream weighing.
Each unresolved contradiction generates a verification hypothesis with specific tool calls and time ranges for Step~2, where segment-specific tools examine targeted audio regions to resolve conflicts.

%\subsection{Answer selection and reasoning generation}
\label{sec:argumentation}

The final stage uses two sequential LLM calls.
An \emph{answer selection} prompt presents the question, formatted audio observations (with source labels but without the source models' answer predictions), reliability evaluations with numeric confidence scores, and tool verification results with confidence and relevance scores; the model selects the best-supported answer.
A subsequent \emph{reasoning generation} call receives the same evidence plus the pre-selected answer and produces prose justification following a seven-section template (what is heard, evidence synthesis, conflict resolution, reliability assessment, tool cross-references, per-choice evaluation, conclusion); a completeness check ensures all observations, tool results, and conflicts are addressed.
%If the output exceeds the character limit, an iterative compaction step preserves key evidence while removing redundancy.
Separating selection from elaboration prevents the model from committing to a narrative before weighing all evidence and lets the reasoning model focus separately on only decision making and then on explanation.

% ===================================================================================
% 4. EXPERIMENTAL SETUP
% ===================================================================================
\section{Experimental setup}
\label{sec:experiments}

\subsection{Dataset and metrics}
\label{sec:dataset}

The Interspeech 2026 Audio Reasoning Challenge uses the 1,000-sample MMAR benchmark~\cite{ma2025mmar}, spanning speech, music, and mixed-modality scenarios across signal, perception, semantic, and cultural reasoning layers.
The challenge evaluates two metrics: answer accuracy and reasoning quality via the MMAR-Rubrics protocol~\cite{ma2026audioreasoningchallenge}.
For each sample, $k{=}5$ checkable criteria are generated from the human-annotated reasoning; an LLM judge assesses each criterion as satisfied or not and the rubrics score is the fraction of satisfied criteria.
Only correct answers receive a reasoning score, while incorrect answers score zero regardless of reasoning quality.
This two-gate design rewards systems that both answer correctly and produce verifiable, evidence-grounded reasoning chains.
The criteria assess factuality, logic, and completeness of the reasoning chain, directly rewarding systems whose output traces each step from evidence to conclusion.

\subsection{Implementation details}
\label{sec:implementation}

The reasoning agent uses moonshotai/Kimi-K2-Thinking~\cite{kimiteam2025kimik2} (temperature 0.6).
LALM evidence comes from stepfun-ai/Step-Audio-R1~\cite{tian2025stepaudior1} and Qwen/Qwen3-Omni-30B-A3B-Thinking~\cite{xu2025qwen3omni}.
ASR uses Whisper large-v3~\cite{radford2023whisper} and Canary~\cite{puvvada2024canary}; source separation uses Demucs~\cite{rouard2023hybrid}; speaker diarization uses pyannote.audio~\cite{bredin2023pyannote}; speaker embeddings use ECAPA-TDNN~\cite{desplanques2020ecapa} via SpeechBrain~\cite{ravanelli2021speechbrain}; pitch estimation uses CREPE~\cite{kim2018crepe}; audio event detection uses PANNs~\cite{kong2020panns}; acoustic features use librosa~\cite{mcfee2015librosa}.

The end-to-end latency of our system averages 8--10 minutes per sample, limiting real-time applicability. 
Outside a competition setting, much of this cost is avoidable: a non-reasoning LLM could likely handle most pipeline stages with minor accuracy loss, and many questions can be answered from speech LLM observations without invoking multiple tool verification loops.

%All non-LLM tools execute in parallel via a Ray GPU cluster.

% ===================================================================================
% 5. RESULTS
% ===================================================================================
\section{Results}
\label{sec:results}

\label{sec:overall}

Our system obtained a reasoning score of 69.8 and 76.9\% accuracy, placing it first on the Agent Track leaderboard (Table~\ref{tab:leaderboard}).

\begin{table}[t]
\caption{Agent Track leaderboard (top 10) and single-model baselines from the MMAR benchmark~\cite{ma2025mmar}. Rubrics = MMAR-Rubrics composite score (ranking criterion); Acc.\ = raw accuracy. Single-model baselines do not have Rubrics scores.}
\label{tab:leaderboard}
\centering
\small
\begin{tabular}{@{}clrr@{}}
\toprule
\textbf{Rank} & \textbf{Team / Model} & \textbf{Rubrics} & \textbf{Acc.} \\
\midrule
\multicolumn{4}{@{}l}{\textit{Agent Track}} \\
1 & TalTech (ours)          & \textbf{69.83} & 76.9\% \\
2 & Team~B        & 66.23 & \textbf{77.4\%} \\
3 & Team~C        & 66.09 & 75.1\% \\
4 & Team~D        & 64.61 & 72.2\% \\
5 & Team~E        & 63.00 & 71.0\% \\
\midrule
\multicolumn{4}{@{}l}{\textit{Commercial and open-weight models (single)}} \\
 & Gemini 2.5 Pro        & --- & 74.7\% \\
 & GPT-4o Audio          & --- & 63.5\% \\
 & Qwen3-Omni-Thinking   & --- & 66.4\% \\
\bottomrule
\end{tabular}
\end{table}

%\subsection{Descriptive analysis}
\label{sec:descriptive}

\begin{table}[t]
\caption{Accuracy by LALM agreement level.}
\label{tab:agreement}
\centering
\small
\begin{tabular}{@{}lrrr@{}}
\toprule
\textbf{Agreement} & \textbf{N} & \textbf{Correct} & \textbf{Accuracy} \\
\midrule
Unanimous   & 128  & 121 & 94.5\% \\
Majority    & 565  & 470 & 83.2\% \\
Conflicting & 307  & 178 & 58.0\% \\
\midrule
Overall     & 1000 & 769 & 76.9\% \\
\bottomrule
\end{tabular}
\end{table}

Descriptive analysis of the submission debug logs revealed the following observations:
\begin{itemize}[nosep,leftmargin=*,parsep=0pt,topsep=0pt]
\item \textbf{Agreement predicts accuracy.}
Table~\ref{tab:agreement} shows a 36.5~pp gap between unanimous (94.5\%) and conflicting (58.0\%) cases, confirming that inter-model agreement is a strong difficulty proxy.
\item\textbf{Confidence is well-calibrated.}
Accuracy increases monotonically with confidence: 91.1\% at $\geq$0.80 ($n{=}237$), 74.4\% at 0.60--0.79 ($n{=}722$), and 39.4\% at 0.40--0.59 ($n{=}33$).
\item \textbf{Corroboration improves accuracy.}
Samples with $\geq$6 corroborated items reach 86.2\% ($n{=}282$) vs.\ 53.8\% with zero ($n{=}13$).
\item \textbf{Tool evidence changes 8.5\% of answers}, overriding both LALM predictions in 85/1,000 cases.
Speech-only questions reach 85.7\% while music drops to 58.7\%.
\item \textbf{Performance varies across reasoning layers.}
Table~\ref{tab:categories} breaks down performance by MMAR question subcategories~\cite{ma2025mmar}.
Semantic tasks are easiest (84.0\%), while cultural reasoning (70.2\%) and signal-level tasks (72.1\%) are hardest.
Music theory (61.9\%), temporal analysis (57.1\%), and audio difference analysis (37.5\%) are the most challenging sub-categories, aligning with the low usefulness of temporal and rhythm tools (Section~\ref{sec:tool_stats}).
\end{itemize}

\begin{table}[t]
\caption{Performance by MMAR reasoning layer and sub-category. \%Max = reasoning quality given correct answers.}
\label{tab:categories}
\centering
%\scriptsize
\setlength{\tabcolsep}{3pt}
\begin{tabular}{@{}cl@{\hspace{4pt}}rrrr@{}}
\toprule
\textbf{Layer} & \textbf{Sub-category} & \textbf{N} & \textbf{Acc.} & \textbf{Reas.} & \textbf{\%Max} \\
\midrule
\multirow{3}{*}{\rotatebox[origin=c]{90}{\textbf{Signal}}}  & Acoustic Quality   & 18  & 88.9 & 84.1 & 94.6 \\
                        & Anomaly Detection  & 17  & 70.6 & 65.5 & 92.8 \\
                        & Audio Difference   & 8   & 37.5 & 25.8 & 68.9 \\
\midrule
\multirow{6}{*}{\rotatebox[origin=c]{90}{\textbf{Perception}}} & Correlation      & 50  & 84.0 & 75.2 & 89.5 \\
                        & Counting \& Stats  & 99  & 62.6 & 54.0 & 86.2 \\
                        & Environ.\ Perc.    & 149 & 82.6 & 75.5 & 91.4 \\
                        & Music Theory       & 63  & 61.9 & 49.6 & 80.2 \\
                        & Spatial            & 15  & 73.3 & 62.2 & 84.9 \\
                        & Temporal           & 28  & 57.1 & 50.5 & 88.3 \\
\midrule
\multirow{3}{*}{\rotatebox[origin=c]{90}{\textbf{Semant.}}}  & Content            & 304 & 83.9 & 76.5 & 91.2 \\
                        & Emotion \& Intent. & 60  & 83.3 & 77.0 & 92.4 \\
                        & Speaker            & 48  & 85.4 & 77.6 & 90.9 \\
\midrule
\multirow{4}{*}{\rotatebox[origin=c]{90}{\textbf{Cultural}}} & Aesthetic Eval.    & 8   & 62.5 & 60.0 & 96.0 \\
                        & Culture of Speaker & 52  & 76.9 & 63.0 & 81.8 \\
                        & Imagination        & 10  & 70.0 & 56.0 & 80.0 \\
                        & Professional Knowl.& 71  & 66.2 & 57.8 & 87.2 \\
\midrule
\multicolumn{2}{@{}l}{\textbf{Overall}} & \textbf{1000} & \textbf{76.9} & \textbf{68.7}\footnotemark & \textbf{89.3} \\
\bottomrule
\end{tabular}
\end{table}
\footnotetext{The reasoning score (68.7) differs from our leaderboard score (69.8) due to running a single evaluation with an LLM judge, whose inherent variability produces minor score fluctuations across runs.}

\subsection{Tool usage and usefulness}
\label{sec:tool_stats}

\begin{table*}[t]
\caption{Per-tool statistics across 1,000 evaluation samples (25 enabled tools, 21 observed), scored by an LLM judge. Sorted by average usefulness score (1--5). Contribution roles: Direct = direct answer, Confirm = confirm hypothesis, Deny = deny hypothesis, Resolve = resolve conflict, Irrel.\ = not useful. Mus.\ = music-specific tool; Int.\ = LLM-interpreted output.}
\label{tab:tool_stats}
\centering
\scriptsize
\resizebox{\textwidth}{!}{%
\begin{tabular}{@{}llccrrrrrrrr@{}}
\toprule
\textbf{Tool} & \textbf{Model / Toolkit} & \textbf{Mus.} & \textbf{Int.} & \textbf{Samples} & \textbf{Avg} & \textbf{Direct} & \textbf{Confirm} & \textbf{Deny} & \textbf{Resolve} & \textbf{Irrel.} & \textbf{\% Irrel.} \\
\midrule
Speech LLM query (Qwen3) & Qwen3-Omni & & & 926 & \textbf{3.51} & 303 & 342 & 78 & 81 & 121 & 13.1\% \\
Transcription & Canary~\cite{puvvada2024canary} / Whisper lg-v3 & & & 602 & \textbf{3.46} & 233 & 181 & 61 & 4 & 122 & 20.3\% \\
Diarization + transcription & Whisper lg-v3 + ECAPA-TDNN~\cite{desplanques2020ecapa} & & & 572 & \textbf{3.33} & 177 & 225 & 20 & 15 & 134 & 23.5\% \\
Speech LLM query (StepAudio) & StepAudioR1 & & & 132 & \textbf{2.87} & 15 & 67 & 11 & 13 & 27 & 20.3\% \\
Melody transcription & CREPE~\cite{kim2018crepe} + librosa & Y & Y & 74 & \textbf{2.74} & 9 & 30 & 7 & 2 & 26 & 35.1\% \\
Instrument detection & PANNs~\cite{kong2020panns} (CNN14) & Y & Y & 213 & \textbf{2.35} & 8 & 92 & 24 & 4 & 85 & 39.9\% \\
Harmonic analysis & librosa (H/P sep., chroma) & Y & & 199 & \textbf{2.25} & 7 & 83 & 9 & 4 & 96 & 48.2\% \\
Beat \& onset detection & librosa (beat\_track, onset) & Y & Y & 104 & \textbf{2.23} & 3 & 41 & 6 & 6 & 48 & 46.2\% \\
Environment detection & librosa (RT60) + heuristics & & & 46 & \textbf{2.04} & 0 & 17 & 3 & 4 & 22 & 47.8\% \\
Synthetic speech detection & AASIST~\cite{jung2022aasist} + prosody & & Y & 5 & \textbf{2.00} & 0 & 0 & 1 & 2 & 2 & 40.0\% \\
Energy dynamics & librosa (RMS, dynamic range) & & & 784 & \textbf{1.99} & 6 & 235 & 58 & 20 & 465 & 59.3\% \\
Spectral features & librosa (centroid, MFCC) & & & 736 & \textbf{1.88} & 4 & 205 & 20 & 8 & 500 & 67.8\% \\
Audio quality & librosa + pyloudnorm (LUFS) & & Y & 17 & \textbf{1.77} & 2 & 3 & 1 & 0 & 11 & 64.7\% \\
Scene context & librosa + PANNs~\cite{kong2020panns} + speech & & & 63 & \textbf{1.44} & 0 & 17 & 1 & 0 & 45 & 71.4\% \\
Chord progression & librosa (chroma\_cqt) + templates & Y & Y & 6 & \textbf{1.33} & 0 & 0 & 1 & 0 & 5 & 83.3\% \\
Speaker count & SpeechBrain~\cite{ravanelli2021speechbrain} ECAPA-TDNN & & & 394 & \textbf{1.26} & 1 & 23 & 12 & 2 & 355 & 90.3\% \\
Event sequence & librosa + PANNs AudioSet & & Y & 264 & \textbf{1.26} & 2 & 22 & 3 & 2 & 235 & 89.0\% \\
Temporal segments & librosa (segment comparison) & & Y & 166 & \textbf{1.22} & 1 & 11 & 3 & 1 & 150 & 90.4\% \\
Audio effects & librosa (reverb, delay, EQ) & & Y & 4 & \textbf{1.25} & 0 & 1 & 0 & 0 & 3 & 75.0\% \\
Tempo tracking & librosa (windowed beat\_track) & Y & Y & 22 & \textbf{1.09} & 0 & 0 & 1 & 0 & 21 & 95.5\% \\
Rhythm analysis & librosa (tempogram) + rules & Y & Y & 8 & \textbf{1.00} & 0 & 0 & 0 & 0 & 8 & 100.0\% \\
\midrule
\multicolumn{12}{@{}l}{\textit{Enabled but not observed during evaluation:}} \\
Source separation & Demucs htdemucs & Y & & 0 & -- & -- & -- & -- & -- & -- & -- \\
Vocal technique & librosa (vibrato, registers) & Y & Y & 0 & -- & -- & -- & -- & -- & -- & -- \\
Instrument sequence & PANNs AudioSet (windowed) & Y & & 0 & -- & -- & -- & -- & -- & -- & -- \\
Rhythm patterns & librosa (tempogram matching) & Y & Y & 0 & -- & -- & -- & -- & -- & -- & -- \\
\bottomrule
\end{tabular}%
}% end resizebox
\end{table*}

To assess per-tool usefulness, an LLM judge (Claude Opus 4~\cite{anthropic2025claude}) independently evaluated each of the 1,000 pipeline debug logs, which contained the full pipeline state (source predictions, tool invocations, reliability evaluations, contradiction traces, and argumentation).
For every tool invocation the judge assigned a usefulness score (1--5) and classified its contribution type (direct answer, confirm/deny hypothesis, resolve conflict, or irrelevant).
Scores should be interpreted as indicative, as judge ratings were not validated against human annotations.

Table~\ref{tab:tool_stats} reports the resulting per-tool statistics across 1,000 samples.
Of 25 enabled tools, 21 were invoked at least once.
LALM queries and ASR tools rank highest in usefulness (avg.\ 3.3--3.5), while temporal and rhythm analysis tools show high irrelevance rates ($>$89\%), suggesting limited utility for the current question distribution.

\subsection{Ablation: Multi-source evidence fusion}
\label{sec:ablation_results}

We run argumentation-level replay ablations using debug logs from the final submission: all upstream artifacts (LALM observations, tool outputs, contradiction traces) are fixed; only the evidence presented to the argumentation model changes.
We use McNemar's test with Holm--Bonferroni correction ($\alpha/2 = 0.025$) to assess statistical significance betweeen the results.

\begin{table}[t]
\caption{Ablation results (argumentation-level replay). $\Delta$ = accuracy change vs.\ baseline in pp; $N_d$ = discordant pairs; $p$ = \textit{p}-value of McNemar' test.}
\label{tab:ablations}
\centering
\begin{tabular}{@{}lrrrr@{}}
\toprule
\textbf{Configuration} & \textbf{Acc.} & $\Delta$ & $N_d$ & $p$ \\
\midrule
Baseline replay                     & 76.6\%  & ---    & ---  & ---     \\
Step-Audio-R1 only                    & 72.3\%  & $-4.3$ & 109  & $<$.001  \\
Qwen3-Omni only                     & 73.4\%  & $-3.2$ & 110  & .003   \\
\bottomrule
\end{tabular}
\end{table}

Table~\ref{tab:ablations} presents replay ablations testing dual- vs.\ single-source evidence.
Both single-source conditions significantly degrade accuracy.

% ===================================================================================
% 6. DISCUSSION
% ===================================================================================
\subsection{Discussion}
\label{sec:discussion}

The ablation results suggest that multi-source fusion gains come not from averaging equally capable models, but from supplementing a strong general-purpose model (Qwen3-Omni) with a specialist that captures complementary observations.
The 85~cases where tool evidence overrides both speech LLMs further demonstrate that acoustic tools provide an independent verification channel.
%The small anchoring effect ($-1.1$~pp, $p = .242$) may reflect the evidence-rich prompt diluting the influence of prediction strings; we retain prediction hiding as a principled precaution~\cite{sharma2024sycophancy}.

%\noindent\textbf{Scope.}
%This paper describes a competition system optimized for the MMAR multiple-choice benchmark under challenge constraints (fixed tool set, fixed evaluation protocol). The reliability weights, confidence caps, and prompt templates were empirically tuned for this setting. Transferring these specific values to other audio reasoning tasks would require re-calibration; however, the architectural pattern of evidence separation and tiered reliability modeling is not benchmark-specific.

The evidence-based architecture is particularly well-suited to the challenge's MMAR-Rubrics evaluation, which scores reasoning chains against instance-specific factuality criteria only when the answer is correct.
Table~\ref{tab:leaderboard} illustrates this alignment: our system places first on Rubrics (69.83) despite slightly lower raw accuracy than Team~B (76.9\% vs.\ 77.4\%).
Three architectural features likely contribute to the reasoning quality advantage noted in Section~\ref{sec:intro}.
First, multi-source evidence gathering produces reasoning dense in citable claims: each observation, tool measurement, and corroboration can satisfy a rubric criterion.
Second, the seven-section reasoning template (Section~\ref{sec:argumentation}) enforces completeness by requiring that every piece of evidence is referenced, conflicts are resolved, and each choice is evaluated.
Third, using a strong reasoning model probably yielded good reasoning scores.
The MMAR-Rubrics evaluation criteria were revealed only after the end of the challenge~\cite{ma2026audioreasoningchallenge}; these architectural choices were not optimized for the rubrics but happen to align with them.

% ===================================================================================
% 7. CONCLUSION
% ===================================================================================
\section{Conclusion}
\label{sec:conclusion}

We described our Interspeech 2026 Audio Reasoning Challenge solution: an ensemble pipeline that manages unreliable audio observations through reliability-tiered evidence combination, dual-source speech LLM fusion, and contradiction-driven verification, placing first on reasoning quality at competitive accuracy on the MMAR benchmark.
Ablations confirm that dual-source fusion provides a statistically significant gain of 3.2--4.3~pp.
Example reasoning outputs are available at \url{https://anonymous.4open.science/r/audio-reasoning-solution-3F43/}.
Future work will investigate scaling beyond multiple-choice settings.

% ===================================================================================
% ACKNOWLEDGMENTS (camera-ready only)
% ===================================================================================
\ifcameraready
\section{Acknowledgments}
This research was supported by the Estonian Centre of Excellence in AI (EXAI), National Program for Estonian Language Technology Program (project EKTB104), both funded by the Estonian Ministry of Education and Research, and by the Estonian Language Data Research Infrastructure (KeTA). Some of the experiments were carried out in the TalTech HPC  \cite{herrmann.btt.m.2025}.
\fi

% ===================================================================================
% GENERATIVE AI DISCLOSURE
% ===================================================================================
\section{Generative AI use disclosure}
Generative AI tools were used for editing and polishing the manuscript and for the post-hoc LLM-as-judge tool evaluation reported in Table~\ref{tab:tool_stats}. All authors take full responsibility for the content of the paper.

% ===================================================================================
% REFERENCES
% ===================================================================================
\bibliographystyle{IEEEtran}
\bibliography{mybib}

\end{document}